# $d^0$ Ferromagnetism in Li-doped ZnO Compounds

L. Chouhan[1], G. Bouzerar[2] and S. K. Srivastava[1*]

[1]Department of Physics, Central Institute of Technology Kokrajhar, Kokrajhar-783370, India
[2]Université Grenoble Alpes, CNRS, Institut NEEL, F-38042 Grenoble, France

[*]**Corresponding Author E-mail:** sk.srivastava@cit.ac.in

## Abstract

Recently, $d^0$ ferromagnetic materials have been projected as one of the promising novel materials for spintronics applications. In this work, we have studied Li-doped ZnO compounds, i.e. $Zn_{1-x}Li_xO$ (x=0, 0.02, 0.04, and 0.06) samples, prepared by the solid-state reaction route method. From the study of crystal structure using X-ray diffraction (XRD) patterns, it is evident that the prepared materials have been formed in a single-phase of the hexagonal wurtzite structure. The refinement of the XRD patterns suggests that there are very small changes in the lattice parameters upon Li-incorporation in ZnO. The average crystallite size ($S_C$), estimated from XRD patterns was found to be in the range of 35-50 nm. The microstructural study by scanning electron microscope reveals the uniform morphology of the grains of the order of 50-70 nm. The energy dispersive spectrum indicates that no unwanted ferromagnetic impurities have crept into the final prepared samples. The measurement of the temperature (T) variation of magnetization (M) with SQUID magnetometer indicates that undoped ZnO exhibits diamagnetic property but all Li-doped compounds exhibit room-temperature ferromagnetism and with a magnetic irreversibility behavior between zero-field cooled and field cooled M-T data. From the magnetization versus field measurements at 3 and 300 K, it is observed that Li-doped samples exhibit ferromagnetic loops with ultra-soft coercivity (~50 Oe) and with a maximum saturation magnetization of 0.10 emu/gm for x= 0.02 sample, which decreases with the increase in Li concentration.

**Key Words:** ZnO; Li-doping; $d^0$ ferromagnetism, Spin Glass Phase

# 1. Introduction:

Recently, doping with non-magnetic elements (e.g. Li, Na, K, Mg, Ag, etc.) in the nonmagnetic host oxides, such as ZnO, $TiO_2$, $SnO_2$, $ZrO_2$ etc. has gained a lot of attention because of their possible application in spintronics [1]. In particular, it was emphasized that ferromagnetism can be observed in such non-magnetic element doped oxide materials and it was referred to as $d^0$ or intrinsic ferromagnetism. There exist a variety of calculations that predicts ferromagnetism even in the un-doped oxide and the observed $d^0$ magnetism was credited to originate from oxygen vacancies/defects [1]. To overcome the difficulties of controlling intrinsic defects, one model was proposed by Bouzerar *et al.*, where it was proposed that nonmagnetic element doping in oxides such as $HfO_2$ or $ZrO_2$ could lead to ferromagnetism (FM) with high Curie temperature [2]. It is important to mention that in the previous past three decades, ferromagnetic ordering in transition metal (TM) doped oxide materials were reported extensively and it was well documented [3-10]. But there is a debate about whether the observed magnetism is an extrinsic effect arising from impurity phases, or it is an intrinsic property [11]. Thus, recently $d^0$ ferromagnetic materials were projected as an alternative pathway to TM doped oxide materials and expected to lead clean materials for spintronics applications. Indeed, many *ab-initio* studies have predicted $d^0$ ferromagnetism in oxides, such as $SnO_2$, $TiO_2$, $ZrO_2$ doped with different non-magnetic elements [12-23] and it was observed experimentally as well [24-33]. Recently, our group has observed $d^0$ ferromagnetism in K-doped $SnO_2$ [27], K-doped rutile $TiO_2$ [28], Li-doped $SnO_2$ [29] and Ag-doped monoclinic $ZrO_2$ compounds [30].

Zinc Oxide (ZnO) has long been used for many applications such as piezoelectric transducers, varistors, and transparent conducting films, etc. Interestingly, $d^0$ ferromagnetism was observed in undoped ZnO thin films also, wherein a successive transition from ferromagnetism to paramagnetism and diamagnetism as a function of film thickness was observed [34-37]. The ferromagnetic order in ZnO thin film was believed to be originated from defects induced [34-37]. However, the ZnO compound prepared in bulk form was found to exhibit diamagnetic behavior [38-39]. In addition to un-doped ZnO compound, the possibility of $d^0$ ferromagnetism in ZnO was explored by doping with various non-magnetic elements, such as Cu-doped ZnO [40-41], alkali metal-doped ZnO [42], Li-doped-ZnO [39, 43-47], Mg-doped ZnO [48] and Ag-doped ZnO [49]. Particularly, the magnetic properties of Li-doped ZnO [39, 43-47] were studied by a few research groups by preparing them in different forms of materials using various methods. Lin *et al.* [43] reported that Li-doped ZnO thin films exhibit paramagnetic behavior; whereas, ferromagnetic behavior was observed in Li-doped ZnO, prepared in the form of nanorods [39], thin-film [45], and nanoparticles using the sol-gel technique [46] and using chemical precipitation method [47]. Chawla *et al.* [39] observed that Li-doped ZnO nanorods exhibit $d^0$ ferromagnetism with high Curie temperature and it was proposed that the substitutional Li dopant induces magnetic moments on neighboring oxygen atoms. Whereas, few other groups [45-46] pointed out that complex defects are responsible for controlling and tuning the observed $d^0$ ferromagnetism in Li-doped ZnO compounds. The maximum saturation magnetization in these compounds was observed to be 0.055 emu/gm. Thus, Li-doped ZnO compounds have led to different results on magnetism and a strong dependency of magnetic property on the method used to prepare and form of the materials was observed.

In a quest of studying the magnetic property of Li-doped ZnO compounds, we have synthesized these compounds in bulk form at equilibrium conditions using the solid-state route method. The Li-doped ZnO bulk compounds, prepared by solid-state route method have the obvious advantage of eliminating the possibility of clustering/defects which is a serious drawback in the materials prepared in thin-film form or prepared by chemical route. We have demonstrated that these Li-doped compounds exhibit room-temperature ferromagnetic behavior.

## 2. Experimental Details

The polycrystalline samples of $Zn_{1-x}Li_xO$ (x=0, 0.02, 0.04, and 0.06) were prepared by solid-state reaction route, using high-purity starting materials of ZnO (Sigma Aldrich, purity-99.999% trace metals basis) and $Li_2CO_3$ (Sigma Aldrich, purity-99.997 % trace metals basis). The maximum amount of any kind of trace magnetic impurities (such as Fe and Co) in the starting materials was found to be less than 2 ppm (parts per million) as mentioned in the supplier chemical analyses report. The required amount of each starting compound (ZnO and $Li_2CO_3$) was calculated in the stoichiometric ratio and the weighing was carried out using an electronic balance. These starting compounds were grinded using a mortar and pestle under acetone for a proper mixing and for good homogeneity of the sample. After grinding the starting materials, the pre-sintering of the samples in powder form was carried out in air at $100^0$C, $200^0$C, and 300˚C for 10 hours at each temperature. The final annealing of these samples in pallet form was done in the air at $500^0$C for 20 hours and these samples were used for the study of crystal structure and magnetic properties. The crystal structure and phase purity of these prepared samples were studied by recording powder X-Ray diffraction (XRD) patterns at room temperature using a Philips diffractometer operated in θ–2θ Bragg-Brentano goniometer geometry and by employing CuK$_α$ (1.541874 Å) incident radiation beam. Data collections were carried out over a scan range of 2θ=20° to angle 2θ=90° with a step of 0.02°. Microstructural study was carried out by recording microstructural images using a ZEISS-ultra-plus scanning electron microscope (SEM). The elemental analysis was done with BRUKER energy dispersive spectrometer (EDS) attached *in-situ* with SEM. The measurements of the temperature variation of magnetization (M-T) under zero-field cooled (ZFC) and field cooled (FC) conditions were carried out under an applied field of 500 Oe and for a temperature range of 3-300 K using commercial SQUID magnetometer (Quantum Design MPMS XL). The field dependence of magnetization (M-H) curves, were measured in the field range of ± 5 T at 3 and 300 K. The magnetic measurements were performed with utmost care and, handling of the samples was done with plastic tweezers.

## 3. Results and Discussion:

To understand the crystal structure and phase purity of all prepared Li-doped ZnO samples, X-Ray diffraction (XRD) patterns were recorded at room temperature and it is presented as Figure 1. Undoped ZnO and Li-doped ZnO samples were found to crystallize in single-phase form with hexagonal ZnO wurtzite structure (space group P6$_3$/mc). The XRD data indicates that no secondary phase is present in the compounds within the instrumental limit of the X-ray diffractometer. In order to further ascertain the phase purity and to estimate the lattice parameters 'a' 'b', 'c' and cell volume (V), the recorded XRD patterns were refined with the help of the Fullprof program software by employing the Rietveld refinement technique [50]. The background coefficient parameters, peak shape parameters (α, β, γ, u, v, w), and lattice parameters (a, b, c) were refined during Rietveld refinement. The XRD patterns for all samples could be refined by using the P6$_3$/mc space group. The XRD patterns along with Rietveld refinement for ZnO and 6% Li-doped ZnO samples are shown in Figure 2. Here the experimental data are shown as open circles and the calculated intensities are shown as solid lines. The bottom line represents the difference between measured and calculated intensities. It is seen that the experimental XRD data matches perfectly with the Rietveld software calculated XRD data. The refinement of the XRD data indicates that the lattice parameters for the un-doped ZnO compound are found to be a=b= 3.2508 Å, c= 5.2076 Å, and V= 47.66 Å$^3$, and these values are comparable with those reported by other groups [40]. The incorporation of Li atoms into ZnO lattice leads to small change in the values of lattice parameters (only in the fourth decimal place), as demonstrated in Figure 3. This can be understood in terms of comparable ionic radii of Zn ions (with ionic radii of 0.74 Å) and Li-ions (with ionic radii of 0.60 Å). However, it is very difficult to comment about whether Li-ions are located at the substitutional sites or interstitial sites. The crystallite size ($S_C$) was calculated from the FWHM (full width at half maximum) data obtained from the refinement and using the Debye Scherrer formula. The crystallite size of un-doped ZnO is estimated to be 35 nm, whereas $S_C$ for 2, 4, and 6 % Li-doped ZnO samples are found to be 37, 49, 38 nm respectively.

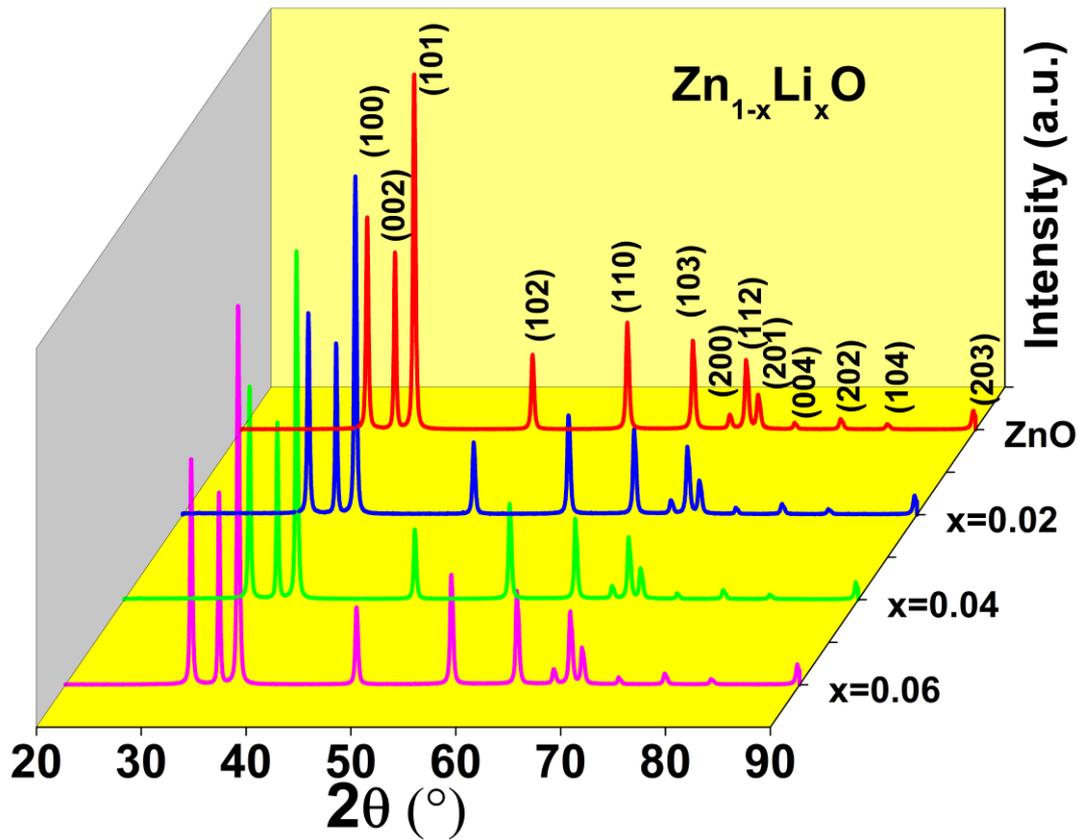

**Figure 1:** X-ray Diffraction patterns of $Zn_{1-x}Li_xO$ (x=0, 0.02, 0.04, and 0.06) compounds.

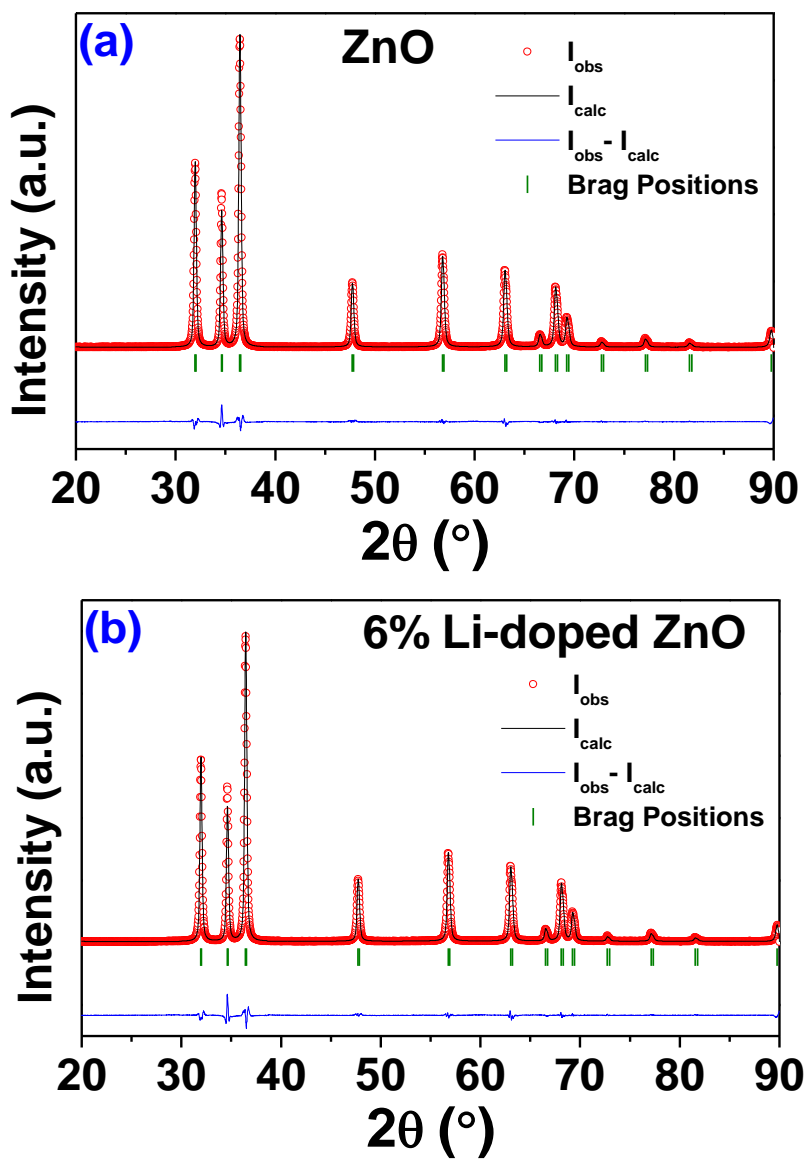

**Figure 2:** XRD patterns along with Rietveld refinement for (a) ZnO (b) 6% Li-doped ZnO samples. The circles represent experimental points and solid line represents Rietveld refined data. The dotted lines show the difference between experimental and refined data.

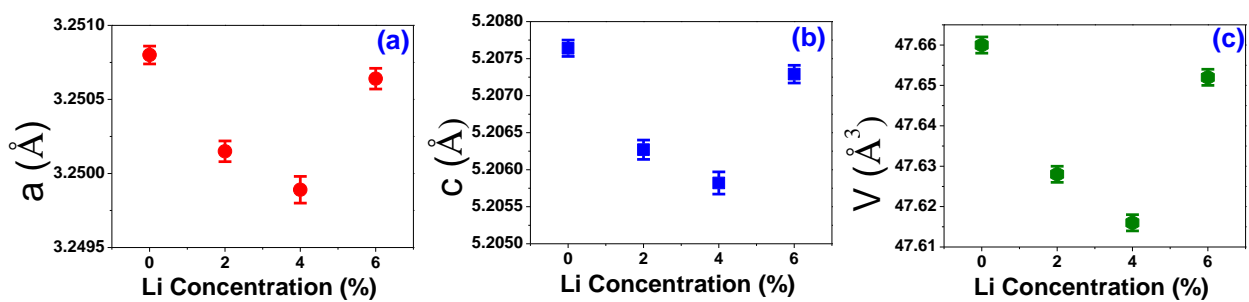

**Figure 3:** Variation of crystal structure parameters a, c, and V, along-with error bar of $Zn_{1-x}Li_xO$ (x=0, 0.02, 0.04 and 0.06) compounds.

Further, from the micro-structural study with SEM, it is observed that the morphology of the samples is quite uniform and nanometric grains have been formed, as demonstrated in Figure 4 (a) and (b) for 2 and 4% Li-doped ZnO samples respectively. The grain size was determined from the analyses of the grain size distribution with software integrated with SEM and; by fitting the distribution of grain size with a Gaussian function. One representative fitting is shown for 4% Li-doped ZnO sample in Figure 4 (c) and the grain size is found of the order of 50-70 nm. In order to ascertain if any other unwanted ferromagnetic impurities (such as Fe or Co etc.) have not crept into the final prepared samples, the elemental analysis of the samples has been carried out by recording EDS spectra. One typical EDS spectrum of 4% Li-doped ZnO sample is shown in Figure 4 (d). Since Li is very light element, it was beyond the instrumental limit of the EDS to detect the distribution of Li. The position for magnetic elements such as Fe and Co is marked as a vertical line, indicating the absence of any magnetic impurity in the sample.

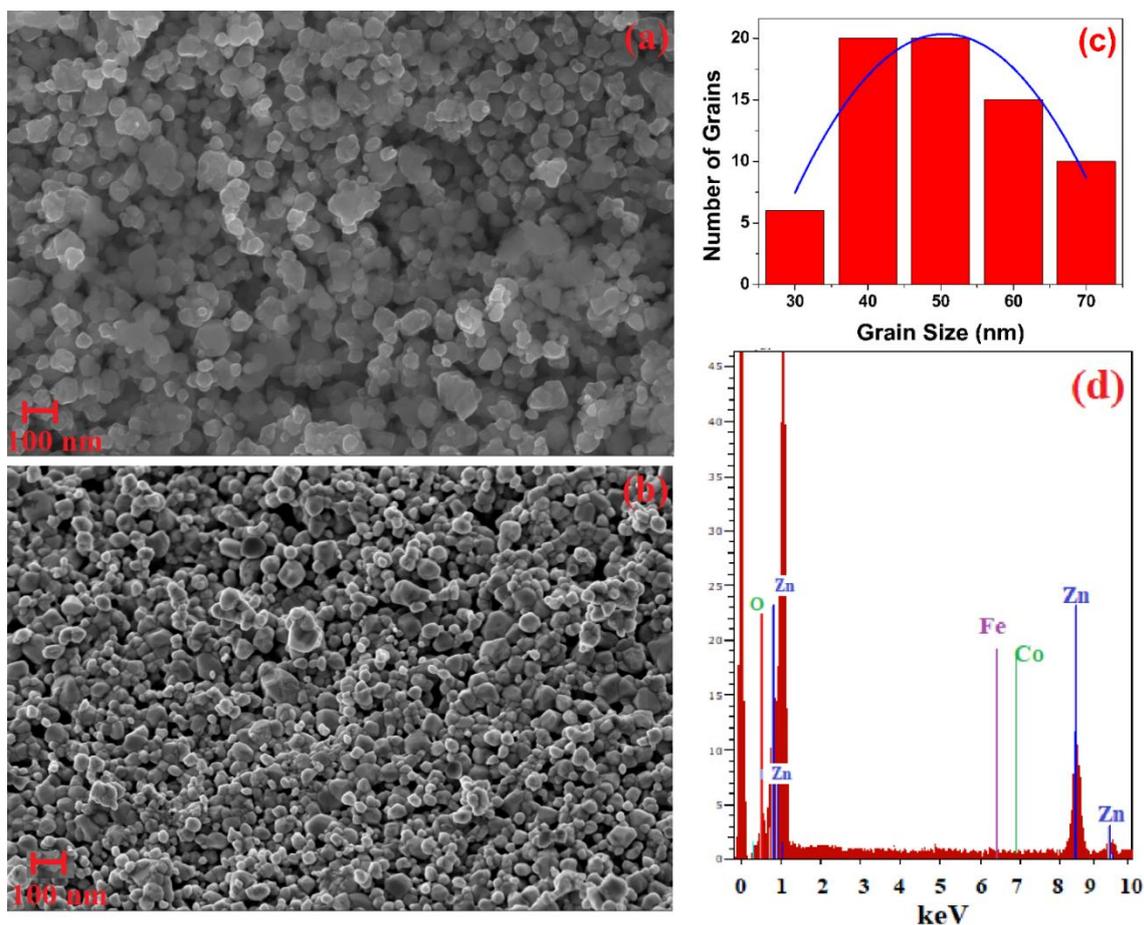

**Figure 4:** SEM images of (a) 2 % and (b) 4 % Li-doped ZnO compounds. (c) Estimation of the mean grain size of 2 % Li-doped ZnO compound by fitting the grain size distribution to the Gaussian curve. (d) EDS spectrum for the 4 % Li-doped ZnO compound. The positions for magnetic elements such as Fe and Co are marked as a vertical line, indicating the absence of any magnetic impurity in the prepared sample.

Followed by the study of crystal structure and micro-structure, the magnetic properties of these prepared samples were studied. As a first step, the magnetic properties of all the starting compounds viz. ZnO and $Li_2CO_3$ have been checked and they clearly exhibit diamagnetic behavior (not shown). The measurements of magnetization versus temperature (M-T) and; field variation of magnetization (M-H) curves (measured at 3 and 300 K) of un-doped ZnO compound indicate a clear diamagnetic behavior, as demonstrated in Figure 5. Similar diamagnetic behavior in the ZnO compound was reported by other groups also [38-39]; but this is unlike other reports on ZnO thin film, where ferromagnetism was observed for the un-doped ZnO compound due to defects [34-37]. The measurement of zero-field cooled (ZFC) M-T curves (Figure 6) for all Li-doped ZnO compounds indicate that there is a very slow dependency of magnetization with temperature throughout the measured temperature range and there was no ferromagnetic transition upto 300 K, indicating that the transition temperature of all Li-doped ZnO compounds is well above the room temperature. Moreover, an upturn/peak in ZFC M-T curves was observed for 2 and 4 % Li-doped ZnO samples at 65 K and 55 K respectively. Here, the characteristic temperature was estimated from fitting the peak with a Gaussian function. We did not observe this feature for 6% Li-doped ZnO sample. This observed feature could originate from small antiferromagnetic phase embedded in the dominant ferromagnetic matrix or due to some disorder in the lattice. The measurement of field cooled (FC) M-T data (as shown in Figure 6) indicates a magnetic irreversibility behavior between ZFC and FC curves for 2, 4 and 6% Li-doped ZnO compounds. The magnetic irreversibility behavior of these samples starts out from 300 K itself and a continuous increase in the value of FC magnetization on lowering the temperature is observed. Moreover, a prominent drastic upturn in the FC magnetization curve, below the characteristic temperature is observed for 2 and 4% Li-doped ZnO samples. Such magnetic irreversibility between ZFC and FC M-T curves may be attributed to the presence of magnetic spin glass phase, possibly arising due to the competing interactions or due to the presence of some disorder in the lattice. A similar kind behavior i.e. presence of magnetic spin glass phase was observed in Li-doped ZnO nanoparticles prepared by chemical sol-gel technique [46].

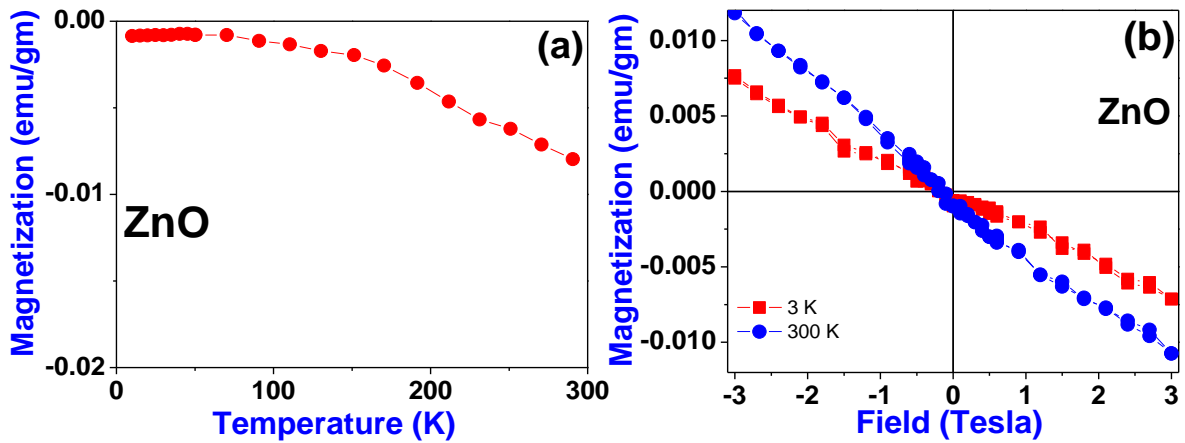

**Figure 5:** (a) Temperature variation of magnetization measured under zero-field cooled condition and an applied field of 500 Oe (b) Magnetization versus field curve measured at 3 and 300 K for ZnO compound.

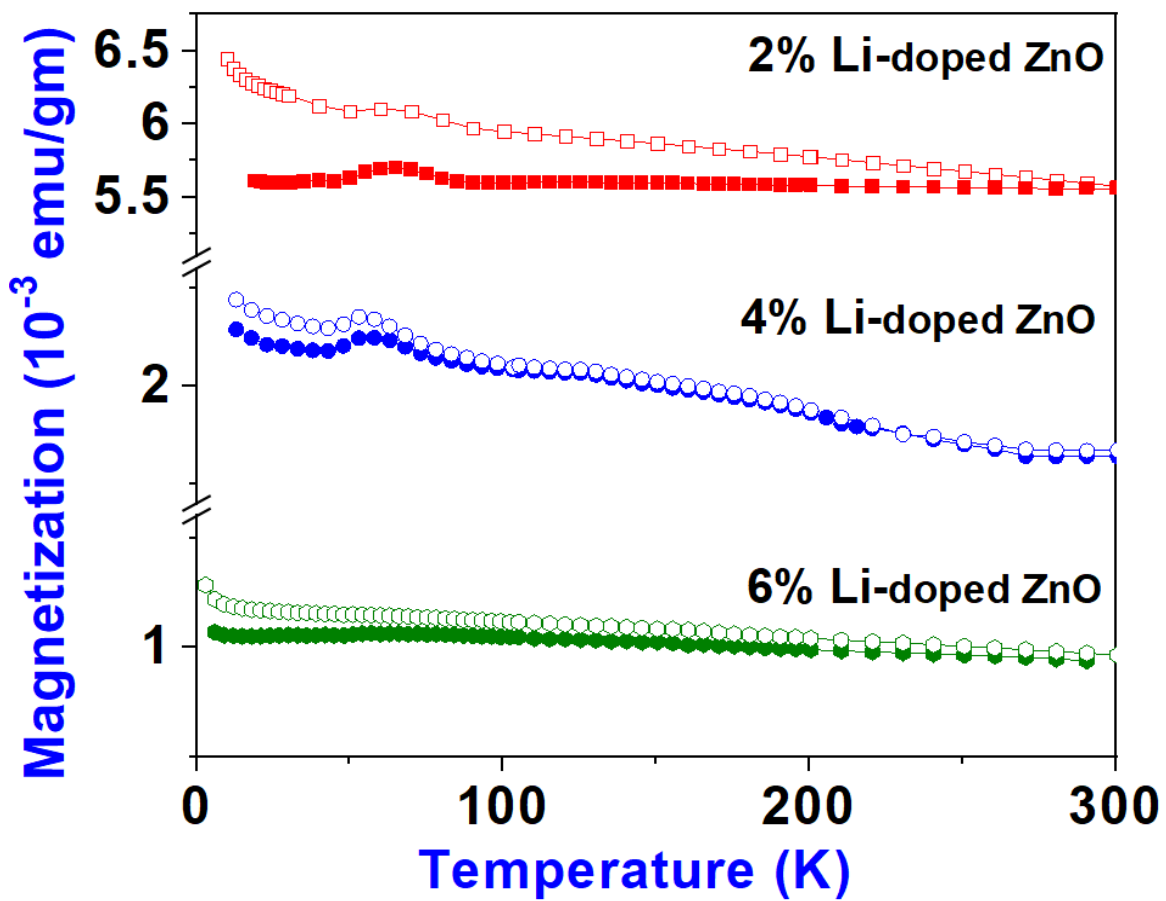

**Figure 6:** Temperature variation of magnetization of 2, 4, and 6% Li-doped ZnO samples measured under an applied field of 500 Oe. The filled symbol represents zero-field cooled M-T and the open symbol represents field-cooled M-T data.
9

To get a further understanding of the magnetic properties of these samples, we have measured field dependence of magnetization in the field range of ± 5 T at 3 K and 300 K for all Li-doped ZnO compounds and they are presented in Figure 7. The M-H curves measured at 3 and 300 K show typical ferromagnetic hysteresis loop for all Li-doped ZnO compounds. The M-H curves measured at 300 K exhibit ultra-soft coercivity, whereas the M-H curves measured at 3 K exhibit enhanced coercivity (as shown in the inset of Figure 7b). The coercivity was found to be for 50, 25 and 48 Oe for 2, 4, and 6% Li-doped samples respectively. These observations indicate that all Li-doped compounds exhibit soft ferromagnetic properties. The value of saturation magnetization ($M_S$) at 300 K (as shown in Figure 8), is observed to be 0.025, 0.008, and 0.002 emu/gm for 2, 4, and 6% Li-doped samples respectively, which is found to be enhanced at 3 K. The value of $M_S$ at 3 K is found to be 0.098, 0.019, and 0.004 emu/gm for 2, 4, and 6% Li-doped samples respectively. Thus, at both the measured temperatures, the value of $M_s$ decreases with the increases of Li concentration. Chawla *et al.* [39] reported the saturation magnetization value of 0.055 emu/gm for 2% Li-doped ZnO, which was the largest $M_S$ value, attributed to substitutional Li. However, Awan *et al.* [41] have demonstrated that the saturation magnetization of 0.055 emu/gm is due to interstitial lithium. Our measured values appear to be higher in comparison to all previously reported values.

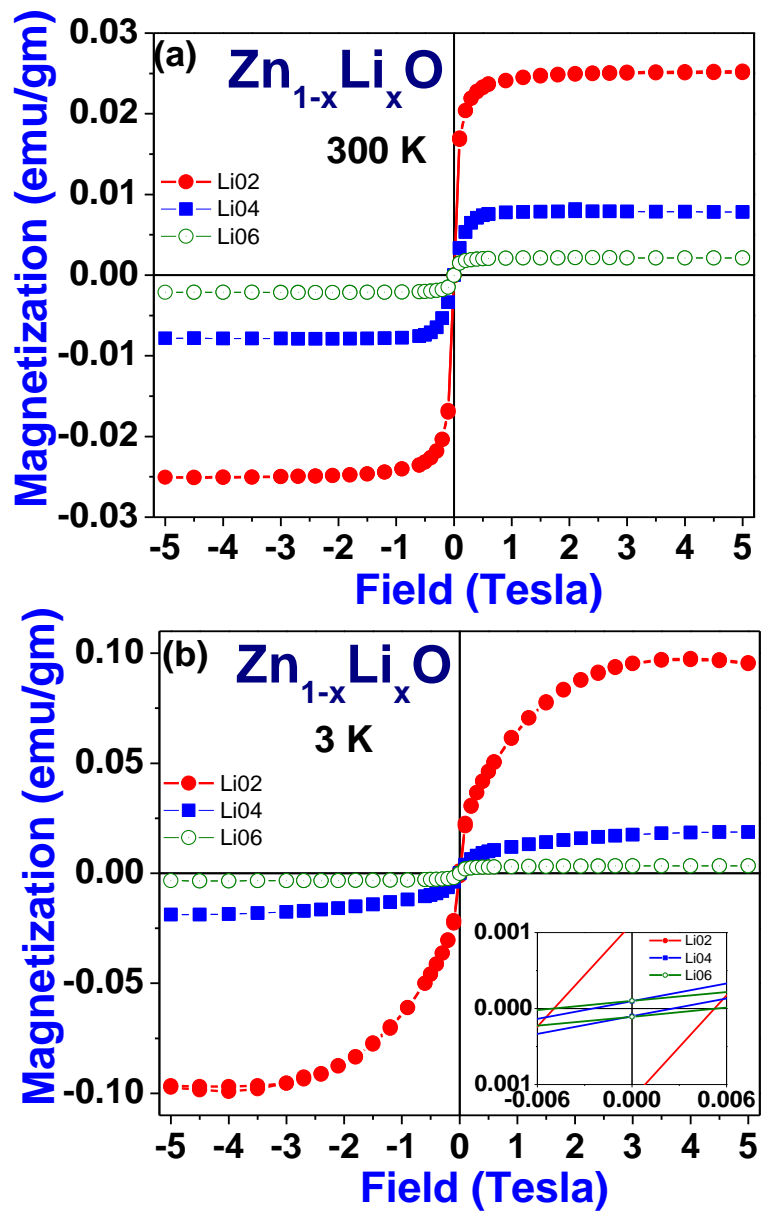

**Figure 7:** Field variation of magnetization for $Zn_{1-x}Li_xO$ (x=0.02, 0.04, and 0.06) compounds measured at (a) 300 K and (b) 3 K. The inset in figure (b) indicates the coercivity of Li-doped ZnO compounds at 3K.

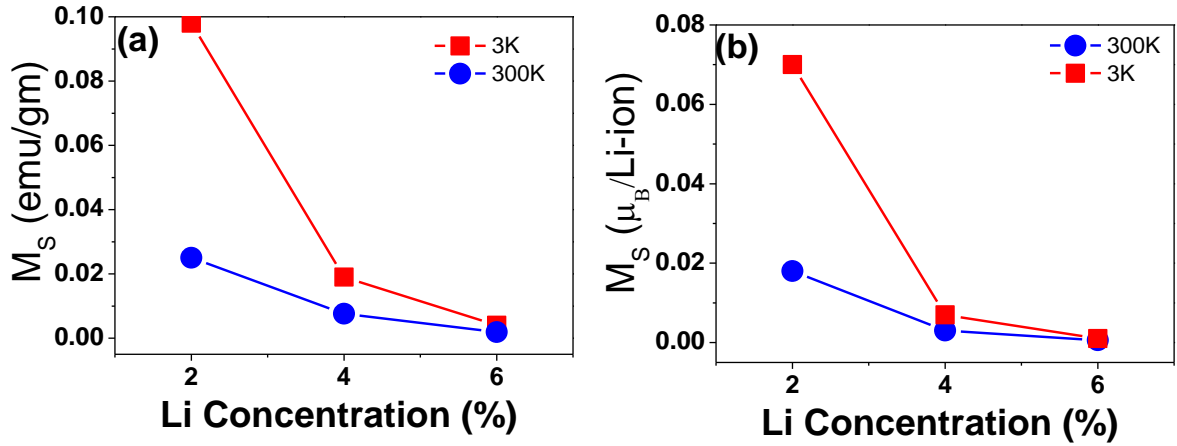

**Figure 8:** The variation of saturation magnetization ($M_S$) with Li concentration, measured at 3 and 300 K for $Zn_{1-x}Li_xO$ (x=0.02, 0.04, and 0.06) compounds.

Now, let us discuss the observed $d^0$ ferromagnetism in the studied Li-doped ZnO compounds. From the crystal structure study, it is observed that the prepared compounds have been formed in pure single-phase and the incorporation of Li atoms into ZnO lattice leads to very small changes in the lattice parameters (only in the fourth decimal place). The magnetic properties, measured from M-T and M-H curves indicate that undoped ZnO compound exhibit diamagnetic behavior, whereas the incorporation of Li into ZnO lattice provoked ferromagnetic behavior with a transition temperature much beyond room temperature. The value of saturation magnetization was found to decreases with the increase in Li concentration. It should be noted that we have not observed any secondary phase from the crystal structure and thus the observed $d^0$ ferromagnetism is intrinsic in nature. In the recent theoretical model of Ref. [2], it was suggested that three physical parameters are essential to explain induced $d^0$ ferromagnetism in the case of a direct non-magnetic cationic substitution in oxide materials: (i) the position of the vacancy (defect) induced impurity level (ii) the density of carrier per defect/vancancy and (ii) the electron-electron interaction strength. It was pointed out that the optimum window of the density of carrier is very important to tune room-temperature ferromagnetism in the non-magnetic element (e.g. Li, Na, K, etc) substitution induced magnetism in oxides. Beyond the optimum window, the magnetic ordering vanishes as magnetic couplings are destroyed by Ruderman-Kittel-Kasya-Yashida (RKKY) type oscillations or antiferromagnetic super-exchange. It was further emphasized by this model that when the on-site potential controlling the impurity level's position is too small, antiferromagnetic nearest-neighbor couplings exist, whereas the coupling become ferromagnetic with the increase of the on-site potential strength. However, with a further increase of this potential, some couplings become antiferromagnetic again. Our experimental results on Li-doped ZnO compounds may fall in line with the predictions of this theory. Our experimental observations show that Li incorporation in ZnO lattice leads to $d^0$ ferromagnetic ordering at room temperature. With the increase of Li doping, a greater number of holes are created and thus it may also create Ruderman-Kittel-Kasuya-Yoshida type oscillations in the couplings, leading to the appearance of spin glass phase, as observed from magnetic irreversibility behavior between ZFC and FC M-T curves. Moreover, RKKY type oscillations in the couplings gradually reduces the magnetic ordering, resulting a decrease of the saturation magnetization with the increase of Li concentration, as observed from the M-H measurement.

## 4. Conclusion

To conclude, we have studied the crystal structure and magnetic properties of Li-doped ZnO compounds, i.e. $Zn_{1-x}Li_xO$ (x=0, 0.02, 0.04 and 0.06) samples, prepared by solid-state reaction route method. From the study of crystal structure by X-ray diffraction (XRD) patterns, it is evident that the prepared materials have been formed in single-phase of the hexagonal wurtzite structure. The refinement of the XRD patterns suggests that there are only small changes in the lattice parameters upon Li-incorporation in ZnO. The average crystallite size ($S_C$) was found to be in the range of 35-50 nm. The microstructural study by scanning electron microscope reveals the uniform morphology of the grains in the range of 50-70 nm. The energy dispersive spectrum indicates that no unwanted ferromagnetic impurities have crept into the final prepared samples. The measurement of the temperature (T) variation of magnetization (M) with SQUID magnetometer indicates that un-doped ZnO exhibits diamagnetic property but all Li-doped compounds exhibit room-temperature ferromagnetism and with a magnetic irreversibility behavior in zero-field cooled and field cooled M-T data. The measurements of hysteresis curves (M-H) at 3 and 300 K indicate that Li-doped samples exhibit ferromagnetic loops with ultra-soft coercivity (~50 Oe) and the maximum saturation magnetization (0.10 emu/gm) was observed for x= 0.02 sample. The value of $M_S$ decreases with an increase in Li concentration. Thus, from the results of M-T and M-H measurements, we can conclusively say that Li-doped compounds exhibit ferromagnetic behavior with a transition temperature much beyond room temperature and co-existence of low-temperature spin-glass phase.

# References:

[1] M. Venkatesan, C.B. Fitzgerald, J.M.D. Coey, Nature 430 (2004) 630.

[2] G. Bouzerar, T. Ziman, Phys. Rev. Lett. 96 (2006) 207602.

[3] H. Ohno, Science 281 (1998) 951.

[4] S. A. Wolf, D. D. Awschalom, R. A. Buhrman, J. M. Daughton, S. V. Molnar, M. L. Roukes, A. Y. Chtcheljanova and D. M. Treger, Science 294 (2001) 1488.

[5] J. K. Furdyna, J. Appl. Phys. 64 (1988) R29.

[6] S.B. Ogale, R.J. Choudhary, J.P. Buban, S.E. Lofland, S.R. Shinde, S.N. Kale, V.N. Kulkarni, J. Higgins, C. Lanci, J.R. Simpson, N.D. Browning, S. Das Sarma, H.D. Drew, R.L. Greene, T. Venkatesan, Phys. Rev. Lett. 91 (2003) 077205.

[7] S. K. Srivastava, P. Lejay, B. Barbara, S. Pailhès, and G. Bouzerar, J. Appl. Phys. 110, (2011) 043929.

[8] S. K. Srivastava, R. Brahma, S. Datta, S. Guha, Aakansha, S. S. Baro, B. Narzary, D. R. Basumatary, M. Kar, S. Ravi, Mater. Res. Express 6 (2019) 126107.

[9] S. K. Srivastava, J Supercond Nov Magn 33 (2020) 2501.

[10] S. K. Srivastava, Aakansha, S. S. Baro, B. Narzary, D. R. Basumatary, R. Brahma, S. Ravi, Journal of Superconductivity and Novel Magnetism 2020 (DOI: https://doi.org/10.1007/s10948-020-05676-y)

[11] T. Dietl, Nature Mater. 9 (2010) 965

[12] W. Zhou, L. Liu, P. Wu, J. Magn. Magn. Mater. 321 (2009) 3356.

[13] C.W. Zhang, S.S. Yan, Appl. Phys. Lett. 95 (2009) 232108.

[14] W.-Z. Xiao, L.-L. Wang, L. Xu, X.-F. Li, H.-Q. Deng, Phys. Status Solidi B 248 (2011) 1961.

[15] J. G. Tao, L. X. Guan, J. S. Pan, C. H. A. Huan, L. Wang, J. L. Kuo,, Phys. Lett. A. 374 (2010) 4451.

[16] F. Maca, J. Kudrnovsky, V. Drchal, G. Bouzerar, Philos. Mag. 88 (2008) 2755.

[17] F. Máca, J. Kudrnovský, V. Drchal, G. Bouzerar, Appl. Phys. Lett. 92 (2008) 212503.

[18] J. Osorio-Guuillén, S. Lany, and A. Zunger, Phys. Rev. Lett. 100 (2008) 036601.

[19] P. Dutta, M.S. Seehra, Y. Zhang, I. Wender, J. Appl. Phys. 103 (2008) 07D104.

[20] M.A. Wahba, S.M. Yakout, J. Sol-Gel Sci. Techn. 92 (2019) 628.

[21] S. Bing, L. Li-Feng, H. De-Dong, W. Yi, L. Xiao-Yan, H. Ru-Qi, K. Jin-Feng, Chin. Phys. Lett. 25 (2008) 2187.

[22] S. Raj, M. Hattori, M. Ozawa, J. Ceram. Soc. Japan 127 (2019) 818. https://doi.org/10.2109/jcersj2.19121.

[23] S. Rani, S. Verma, S. Kumar, Appl. Phys. A 123 (2017) 539.

[24] D. L. Hou, H. J. Meng, L. Y Jia, X. J. Ye, H. J. Zhou, X. L. Li, EPL 78 (2007) 67001.

[25] S. Duhalde, M.F. Vignolo, F. Golmar, C. Chiliotte, C.E.R. Torres, L.A. Errico, A.F. Cabrera, M. Rentería, F.H. Sánchez, M. Weissmann, Phys. Rev. B 72 (2005) 043705.

[26] X.J. Ye, W. Zhong, M.H. Xu, X.S. Qi, C.T. Au, Y.W. Du, Phys. Lett. A 373 (2009) 3684.

[27] S. K. Srivastava, P. Lejay, B. Barbara, S. Pailhès, V. Madigou, and G. Bouzerar, Phys. Rev. B 82 (2010) 193203.

[28] S. K. Srivastava, P. Lejay, A. Hadj-Azzem, G. Bouzerar, J. Supercond. Nov. Magn. 27 (2013) 487.

[29] S. K. Srivastava, P. Lejay, B. Barbara, O. Boisron, S. Pailhès and G. Bouzerar, J. Phys.: Condens. Matter 23 (2011) 442202.

[30] J. Wang, D. Zhou, Y. Li, P. Wu, Vacuum, 141 (2017) 62.

[31] M. C. Dimri, H. Khanduri, H. Kooskora, M. Kodu, R. Jaaniso, I. Heinmaa, A. Mere, J. Krustok, R. Stern, J. Phys. D: Appl. Phys. 45 (2012) 475003.

[32] S. Akbar, S. K. Hasanain, O. Ivashenko, M. V. Dutka, N. Z. Ali, G. R. Blake, J. Th. M. De Hosson and P. Rudolf, RSC Adv., 10 (2020) 26342.

[33] L. Chouhan, G. Bouzerar, S. K. Srivastava, Vacuum 182 (2020) 109716.

[34] M. Kapilashrami, J. Xu, V. Ström, K. V. Rao, and L. Belova, Appl. Phys. Lett. **95**, (2009) 033104.

[35] S. Mal, J. Narayan, S. Nori, J. T. Prater, D. Kumar, Solid State Communications, 150 (2010) 1660.

[36] S. Mal, T-H Yang, C. Jin, S. Nori, J. Narayan, J. T. Prater, Scripta Materialia 65 (2011) 1061.

[37] S. Mal, S. Nori, C. Jin, J. Narayan, S. Nellutla, A. I. Smirnov, and J. T. Prater Journal of Applied Physics 108 (2010) 073510.

[38] G. Vijayaprasath, R. Murugan Y. Hayakawa, G. Ravi, Journal of Luminescence 178 (2016) 375.

[39] S. Chawla, K. Jayanthi, R.K. Kotnala, Phys. Rev. B 79 (2009) 125204.

[40] D. Chakraborti and J. Narayan, Appl. Phys. Lett. 90(2007) 062504.

[41] N. Ali, B. Singh, Z.A. Khan, V. A. R., K. Tarafder, S. Ghosh, *Sci Rep* 9 (2019) 2461.

[42] S. Chawla, K. Jayanthi, R.K. Kotnala, J. Appl. Phys. 106 (2009) 113923.

[43] Y.-H. Lin, M. Ying, M. Li, X. Wang, and C.-W. Nan, Appl. Phys. Lett. 90 (2007) 222110.

[44] R. Vettumperumal, S. Kalyanarama, B. Santoshkuma, R. Thangavel, Materials Research Bulletin, 50 (2014) 7.

[45] J.B. Yi, C.C. Lim, G.Z. Xing, H.M. Fan, L.H. Van, S.L. Huang, K.S. Yang, X.L. Huang, X.B. Qin, B.Y. Wang, T. Wu, L. Wang, H.T. Zhang, X.Y. Gao, T. Liu, A.T.S. Wee, Y.P. Feng, J. Ding, Phys. Rev. Lett. 104 (2010) 033501.

[46] S. U. Awan, S. K. Hasanain, M. F. Bertino, and G. H. Jaffari, J. Appl. Phys. 112 (2012) 103924.

[47] B. K. Pandey, A. K. Shahi, J. Sha, R. K. Kotnala, R. Gopal, Journal of Alloys and Compounds, 823 (2020) 153710.

[48] P. Kumar, Y. Kumar, H.K. Malik, *et al. Appl. Phys. A* 114 (2014) 453.

[49] N. Ali, Vijaya A. R., Z. A. Khan, K. Tarafder, A. Kumar, M. K. Wadhwa, B. Singh & S. Ghosh, Scientific Reports 9 *(2019) 20039.*

[50] R. A. Young, The Rietveld Method (International Union of Crystallography), reprint first ed., Oxford University Press, New York, 1996.

[51] M. A. Peck, Y. Huh, R. Skomski, R. Zhang, P. Kharel et al., J. Appl. Phys. 109 (2011) 07B518.